\documentclass{eas}
\usepackage{graphicx}

\begin{document}

\title{Planetary nebulae: What can they tell us about close binary evolution?} 
\runningtitle{PNe: What can they tell us about close binary evolution?}
\author{David Jones}\address{Instituto de Astrof\'isica de Canarias, E-38200 La Laguna, Tenerife, Spain; \email{djones@iac.es}}\secondaddress{Departamento de Astrof\'isica, Universidad de La Laguna, E-38206 La Laguna, Tenerife, Spain}
%
%
\begin{abstract}
It is now clear that a binary pathway is responsible for a significant fraction of planetary nebulae, and the continually increasing  sample of known central binaries means that we are now in a position to begin to use these systems to further our understanding of binary evolution.  Binary central stars of planetary nebulae are key laboratories in understanding the formation processes of a wide-range of astrophysical phenomena - a point well-illustrated by the fact that the only known double-degenerate, super-Chandrasekhar mass binary which will merge in less than a Hubble time is found inside a planetary nebula.  Here, I briefly outline our current understanding and avenues for future investigation.
\end{abstract}
\maketitle
\section{Introduction}

Central star binarity has long been invoked as playing an important role in the shaping of aspherical planetary nebulae (PNe), while several works have shown that binary interaction may, in fact, be the only feasible source of asphericity in PNe (Nordhaus \& Blackman \cite{nordhaus06}; Garc\'ia-Segura {\em et al.\/} \cite{garcia-segura14}).  Until very recently, only a handful of binary central stars (bCSPN) were known, however the advent of wide-field variability surveys led to a doubling of number known (see e.g.\ Miszalski {\em et al.\/} \cite{miszalski09a}).  As a result, the sample was sufficient to begin to identify morphological features which appear more prevalent amongst the population of close binaries than the general PN population, and to begin to use these as selectors for targeted searches (Miszalski {\em et al.\/} \cite{miszalski11c}).  The use of these selectors (namely jets, filaments \& rings) has led to a further increased rate of discovery led by the author and collaborators (e.g.\ Corradi {\em et al.\/} \cite{corradi11}; Miszalski {\em et al.\/} \cite{miszalski11a,miszalski11b}; Boffin {\em et al.\/} \cite{boffin12}; Jones {\em et al.\/} \cite{jones14,jones15}). 

\begin{figure}
\centering
\includegraphics[angle=270,width=0.7\textwidth]{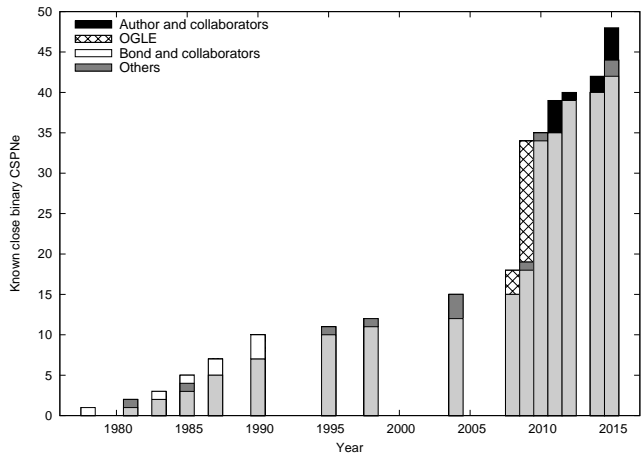}
\caption{The discovery rate of close binary CSPN, updated from Miszalski {\em et al.\/} (\cite{miszalski11c}).  For full references see the list maintained at http://drdjones.net/bCSPN.}
\label{fig:discoveries}
\end{figure}

Now, as we approach a statistically significant sample of close binary central stars ($\sim$50, figure \ref{fig:discoveries}), it is important to begin to look for further connections between these systems - not only to understand the role binarity plays in the formation of planetary nebulae, but also what these systems can tell us about binary evolution in general.  In these proceedings, I will attempt to outline the importance of planetary nebulae in understanding binary evolution as well as what we have already learnt from them.

\section{Supernovae type Ia}

Recent works have estimated that at least 20\% supernovae type Ia (SNe Ia) progenitors will have previously passed through a PN phase (Tsebrenko \& Soker \cite{tsebrenko15}) irrespective of the mechanism responsible for the SN explosion.  One of the favoured mechanisms is the merger of a double-degenerate close binary system where the total mass exceeds the Chandrasekhar mass.  Several such systems are known, but only one is close enough to merge in less than a Hubble time\footnote{There are, however, other candidate systems where the total mass is uncertain.} - the central binary of the planetary nebulae Hen~2-428 (Santander-Garc\'ia {\em et al.\/} \cite{santander-garcia15}).  Furthermore, it is important to note that the fraction of double-degenerates amongst known bCSPNe is rather high (perhaps $\sim$20\% based on variability studies), given that these systems are generally very difficult to detect via photometric monitoring (De Marco \cite{demarco09}). This further strengthens the connection between PNe and understanding the formation of SNe Ia from mergers in double-degenerate systems.

\section{Common-envelope evolution}

All of the close binaries shown in figure \ref{fig:discoveries} have passed through a common-envelope (CE) phase of evolution.  The CE phase is critical in the formation of a wide-range of astrophysical phenomena (including but not limited to cataclysmic variables, low mass X-ray binaries, novae and, of course, SN Ia).  In spite of its importance, the CE is still very poorly understood, with hydrodynamic simulations failing to eject the entire envelope (Passy {\em et al.\/} \cite{passy12}) and a reliance on the intrinsically flawed $\alpha$ formalism (De Marco \cite{demarco09}).  PNe with binary central stars can provide valuable data for understanding the CE given that the nebula itself is assumed to be the ejected CE, and that the central binary is ``fresh-out-of-the-oven'' and yet to have time to adjust.  It is, therefore, interesting to note that in all cases of a main sequence dwarf secondary, where the parameters of the binary system are well constrained (masses, temperature, radii, etc.), the secondaries are found to be ``inflated'' (Jones {\em et al.\/} \cite{jones15}).  This inflation is almost certainly due to a phase of rapid mass transfer immediately prior to the CE phase (which has thrown the secondary out of thermal equilibrium).  This hypothesis is further supported by the presence of jets (a product of mass transfer) which are older than the central nebular regions (i.e.\ the ejected envelope, Corradi {\em et al.\/} \cite{corradi11}; Boffin {\em et al.\/} \cite{boffin12}), and in the case of The Necklace a secondary which has been contaminated with chemically enriched material (Miszalski, Boffin \& Corradi \cite{miszalski13}).  This pre-CE mass transfer is a critical constraint for future modelling of the CE process.

\section{Unusual chemistry}

As well as the aforementioned morphological similarities amongst PNe known to host close binary central stars, recent work has highlighted a connection to unusually high abundance discrepancy factors (adfs, Corradi {\em et al.\/} \cite{corradi15}).  It is a well known problem in nebular astrophysics that abundances derived from optical recombination lines exceed those derived using collisionally excited lines by a factor of a few and occasionally more, with this ratio being known as the adf.  Most PNe are found to display adfs from 1--5, however a small fraction show extremely elevated adfs reaching well above 100 - and almost all of these elevated adfs are found in systems with a binary central star (or suspected binary central star).  These high adfs are believed to arise from a two phase gas, whereby one phase has a typical temperature (10kK) and composition while the second has a much lower temperature and is metal enriched (Liu {\em et al.\/} \cite{liu06}), something which might naturally arise from a binary evolution.

\section{Summary}

Binary interaction is now the preferred mechanism by which asphericities develop in PNe, but with the growing number of known bCSPNe (and a similar increase in number for which parameters have been derived through detailed study) it is now possible to use these systems to try and resolve key problems in our understanding of binary evolution - namely the CE phase.  PNe indicate that there may be a phase of intense mass transfer immediately prior to entering the CE, something which may prove invaluable in producing models which reproduce the observed parameters of post-CE systems.  Furthermore, the only known double-degenerate super-Chandrasekhar mass system which will merge in less than a Hubble time is found inside a planetary nebula, showing that PNe are key to understanding a wide variety of other phenomena, including the mechanism by which SN Ia occur.  An intriguing connection between high adfs and central star binarity has recently been highlighted.  Understanding how these adfs arise will almost certainly provide further insight into close binary evolution.


\end{document}